\documentclass[aps, pra, twocolumn,notitlepage, amsmath, amssymb, superscriptaddress, nofootinbib]{revtex4-1}

\makeatletter
\def\p@subsection{}
\def\p@subsubsection{}
\makeatother

\usepackage[utf8]{inputenc}
\usepackage{graphicx}
\usepackage{units}
\usepackage{color}
\usepackage[pdftex, colorlinks=true, linkcolor=myblue, citecolor=myblue,
urlcolor=myblue]{hyperref}
\usepackage{times}
\usepackage{comment}
\usepackage{graphicx}
\usepackage{amsmath, calc}
\usepackage{mathrsfs}
\usepackage{MnSymbol}
\usepackage{amsfonts}
\usepackage{amssymb}
\usepackage{bm}
\usepackage{bbm}
\usepackage{color}
\usepackage{array}
\usepackage{units}
\usepackage{mathrsfs,amsmath}

\definecolor{myblue}{rgb}{0,0,1}
\definecolor{myred}{rgb}{1,0,0}

\newcommand{\expect}[1]{\langle #1 \rangle}

\DeclareMathOperator{\Tr}{Tr}

\begin{document}
%\narrowtext
\title{Exceptional points in oligomer chains}

\author{Charles~Andrew~Downing}
\affiliation{Department of Physics and Astronomy, University of Exeter, Exeter EX4 4QL, United Kingdom}

\author{Vasil~Arkadievich~Saroka}
\affiliation{Center for Quantum Spintronics, Department of Physics, Norwegian University of Science and Technology, NO-7491 Trondheim, Norway}
\affiliation{Institute for Nuclear Problems, Belarusian State University, Bobruiskaya 11, 220030 Minsk, Belarus
\\ email:~c.a.downing@exeter.ac.uk,~vasil.saroka@ntnu.no}

\date{\today}
%===========================================================================
%===========================================================================
%===========================================================================
%===========================================================================
%===========================================================================
%===========================================================================
%===========================================================================
%===========================================================================

\begin{abstract}
\noindent \textbf{Abstract}\\
Symmetry underpins our understanding of physical law. Open systems, those in contact with their environment, can provide a platform to explore parity-time symmetry. While classical parity-time symmetric systems have received a lot of attention, especially because of the associated advances in the generation and control of light, there is much more to be discovered about their quantum counterparts. Here we provide a quantum theory which describes the non-Hermitian physics of chains of coupled modes, which has applications across optics and photonics. We elucidate the origin of the exceptional points which govern the parity-time symmetry, survey their signatures in quantum transport, study their influence for correlations, and account for long-range interactions. We also find how the locations of the exceptional points evolve as a function of the chain length and chain parity, capturing how an arbitrary oligomer chain transitions from its unbroken to broken symmetric phase. Our general results provide perspectives for the experimental detection of parity-time symmetric phases in one-dimensional arrays of quantum objects, with consequences for light transport and its degree of coherence.
\end{abstract}

%===========================================================================
%===========================================================================
%===========================================================================
%===========================================================================
%===========================================================================
%===========================================================================
%===========================================================================
%===========================================================================

\maketitle

%===========================================================================
%===========================================================================
%===========================================================================
%===========================================================================
%===========================================================================
%===========================================================================
%===========================================================================
%===========================================================================

\noindent \textbf{Introduction}\\
While the eigenvalues of a Hermitian Hamiltonian are always real, the Hermicity condition is more stringent than is strictly necessary~\cite{Mostafazadeh2008}. It was shown by Bender and co-workers that Hamiltonians that obey parity-time ($\mathcal{PT}$) symmetry can both admit real eigenvalues and describe physical systems~\cite{Bender1998, Bender2018}. The condition of combined space and time reflection symmetry has immediate utility for some open systems, where there is balanced loss into and gain from the surrounding environment. The application of the concept of $\mathcal{PT}$ symmetry into both classical and quantum physics has already led to some remarkable advances and unconventional phenomena, which cannot be captured with standard Hermitian Hamiltonians~\cite{Konotop2016, Martinez2018,Foa2020,Bergholtz2020,Ashida2020}.

An important concept within $\mathcal{PT}$ symmetry is that of exceptional points. Let us consider the simplest case of a pair of coupled oscillators, each of resonance frequency $\omega_0$ and interacting via the coupling constant $g$. The two resulting eigenfrequencies $\omega_2$ and $\omega_1$ are given by $\omega_{2, 1} = \omega_0 \pm g$. After including gain at a rate $\kappa$ into the first oscillator and an equivalent loss $\kappa$ out of the second oscillator, the renormalized eigenfrequencies $\omega_{2}'$ and $\omega_{1}'$ of this $\mathcal{PT}$-symmetric setup become $\omega_{2, 1}' = \omega_0 \pm \sqrt{g^2 - \kappa^2/4}$ [see Supplementary Note 1]. The exceptional point (for this $\mathcal{N} = 2$ oscillator system) is 
 \begin{equation}
\label{eqapp:ffsdssdsddsdfsg22}
 \left( \frac{g}{\kappa} \right)_{\mathcal{N} = 2} = \frac{1}{2},
 \end{equation}
which defines the crossover between the unbroken $\mathcal{PT}$ phase with wholly real $\omega_{2, 1}'$, and the broken phase with complex $\omega_{2, 1}'$. Therefore, by modulating the ratio $g/\kappa$ one can induce a plethora of (sometimes unexpected) phenomena intrinsically linked to $\mathcal{PT}$ symmetry, for example in light transport where amplification and attenuation readily arise~\cite{Konotop2016, Martinez2018, Foa2020, Bergholtz2020, Ashida2020}. 

\begin{figure}[tb]
 \includegraphics[width=1.0\linewidth]{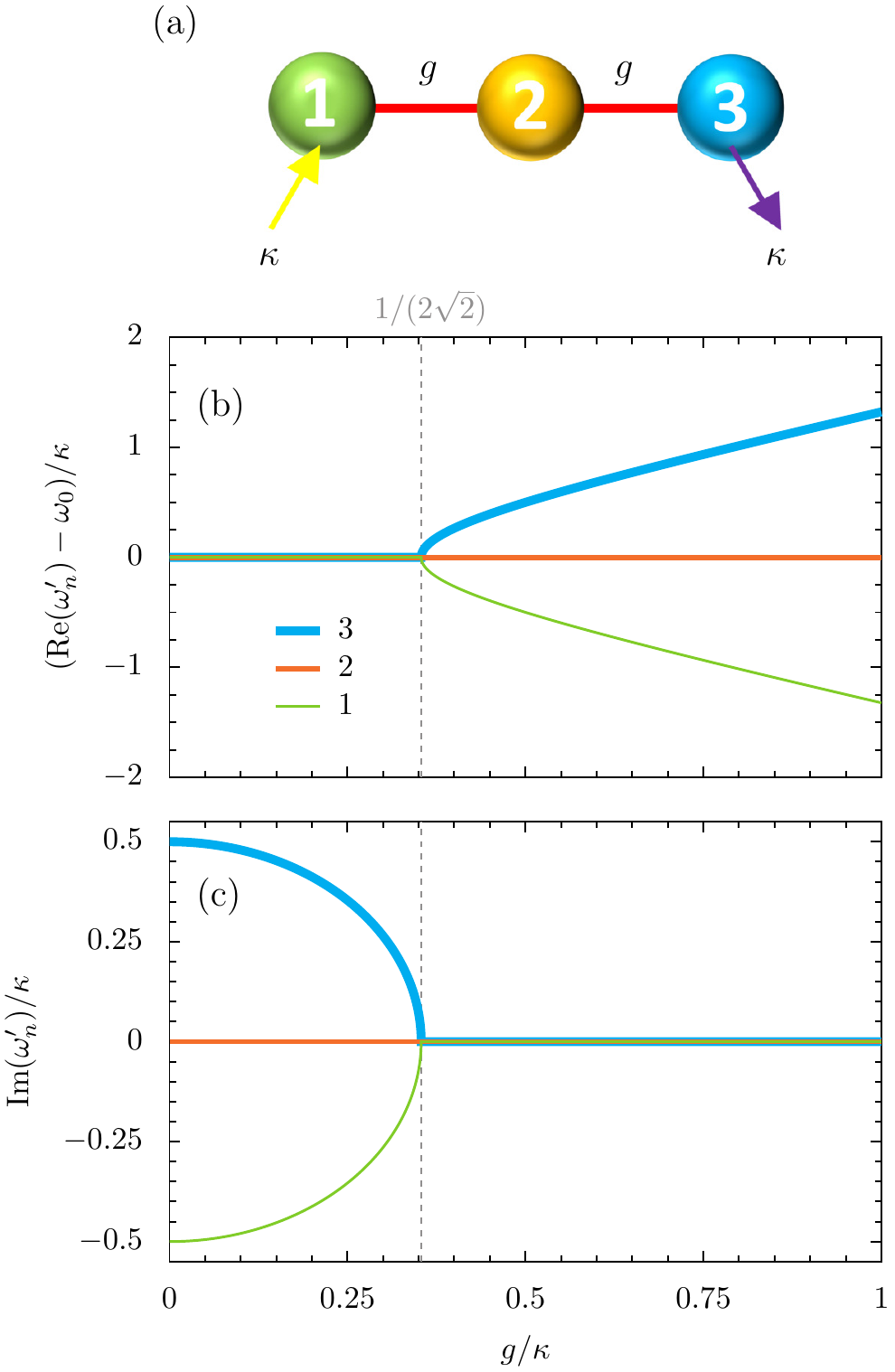}
 \caption{ \textbf{The $\mathcal{PT}$-symmetric trimer.} Panel (a): a sketch of the $\mathcal{PT}$-symmetric trimer chain (colored spheres), where each oscillator is of resonance frequency $\omega_0$, and the coupling constant is $g$. The first oscillator (green sphere) is subject to gain $\kappa$ (yellow arrow), and the final oscillator (cyan sphere) to loss $\kappa$ (purple arrow). Panel (b): the real parts of the eigenfrequencies $\omega_n'$ of the trimer, measured from $\omega_0$, as a function of $g$, in units of $\kappa$ [Eq.~\eqref{eq:Eig_one_PT}]. Panel (c): the corresponding imaginary parts of the eigenfrequencies $\omega_n'$. Vertical dashed lines: exceptional points marking broken-unbroken $\mathcal{PT}$ symmetry phases [Eq.~\eqref{eqapp:ffsdsdsdfsg22}]. }
 \label{exceptional}
\end{figure}

Recently, optical and photonic systems have become popular playgrounds to test $\mathcal{PT}$-symmetric effects in the laboratory~\cite{Ganainy2018, Ozdemir2019, Miri2019}. Indeed, recent experiments in the area have seen non-reciprocal light propagation in coupled waveguides~\cite{Ruter2010}, single-mode lasing in microring cavities~\cite{Feng2014}, extraordinary transmission in microtoroidal whispering-gallery-mode resonators~\cite{Peng2014}, and the development of hybrid optoelectronic devices~\cite{Liu2018}. In parallel, there has been much theoretical work on many-mode $\mathcal{PT}$-symmetric systems~\cite{Suchkov2016b}, including considerations of trimers~\cite{Li2013a, Downing2020a}, quadrimers~\cite{Zezyulin2012, Gupta2014, Rivolta2015, Gupta2015, Ding2016, Kalozoumis2016, Rivolta2016, Liu2017, Jouybari2017, Zhou2018, Kalaga2019} and more generally oligomers~\cite{Jin2009,Jin2010, Joglekar2010,Joglekar2011,Li2011, Ambroise2012, Duanmu2013,Liang2014,Jin2017,Ortega2020,Flynn2020,Arkhipov2020,Arkhipov2021,Nakagawa2021,Agarwal2021,McDonald2021}.

Inspired by the pioneering experiments of Hodaei and co-workers with chains of ring-shaped optical resonators~\cite{Hodaei2017}, we develop a simple theory of short oligomer chains in an open quantum systems approach. In particular, we study dimer ($\mathcal{N} = 2$), trimer ($\mathcal{N} = 3$) and quadrimer ($\mathcal{N} = 4$) chains in detail~[see also Supplementary Notes 1 and 2]. We derive the locations of the exceptional points, and explore the influence of the $\mathcal{PT}$ symmetry phase on both the population dynamics (revealing regions of amplification) and for correlations (showing areas of perfect coherence and incoherence). Our open quantum systems approach follows in the wake of a number of recent theoretical works~\cite{Minganti2019, Hatano2019,Minganti2020,Wiersig2020, Huber2020, Jaramillo2020,Purkayastha2020,Kumar2021} which employ the concept of $\mathcal{PT}$ symmetry with quantum master equations. We note that a related and pioneering experiment with superconducting qubits has latterly been reported~\cite{Chen2021}, highlighting the timeliness of quantum $\mathcal{PT}$-symmetry. We also uncover how the exceptional point of Eq.~\eqref{eqapp:ffsdssdsddsdfsg22} is generalized for an oligomer chain of an arbitrary size $\mathcal{N}$, where there is gain into the first oscillator and an equivalent loss out of the last oscillator, with neutral oscillators in between. Using a transfer matrices approach, we derive an interesting scaling with $\mathcal{N}$ of $(g/\kappa)_{\mathcal{N}}$, and we find a feature due to the parity of the oligomer which provides tantalizing opportunities for experimental detection. Finally, we investigate the emergent and rich $\mathcal{PT}$ symmetry phase diagrams when long-range coupling (beyond nearest-neighbor) is taken into account, which crucially determines whether the exceptional point is of higher order (compared to the dimer case) or not.\\

%===========================================================================
%===========================================================================
%===========================================================================
%===========================================================================
%===========================================================================
%===========================================================================
%===========================================================================
%===========================================================================

\noindent \textbf{Results and Discussion}\\
\noindent \textbf{\textit{Trimer chain: model}}\\
Here we look at simplest nontrivial chain of oscillators, the trimer chain ($\mathcal{N} = 3$ sites). The trimer [sketched in Fig.~\ref{exceptional}~(a)] already displays some interesting phenomena which is common across all odd-sited oligomers, and yet it retains some beauty due to its simplicity. The Hamiltonian operator for the trimer chain reads (we set $\hbar = 1$ throughout)
\begin{equation}
\label{eq:Ham}
\hat{H} = ~\omega_0 \left( b_{1}^{\dagger} b_{1}  + b_{2}^{\dagger} b_{2} + b_{3}^{\dagger} b_{3} \right) + g \Big( b_{1}^{\dagger} b_{2} +  b_{2}^{\dagger} b_{3} + \mathrm{h.c.} \Big),
\end{equation}
where the bosonic creation (annihilation) operators on oscillator $n$ is denoted by $b_n^\dagger$ ($b_n$). Each oscillator is associated with the resonance frequency $\omega_0$, and the nearest-neighbor coupling between sites is given by $g$. The three eigenfrequencies $\omega_n$ of the trimer read
\begin{subequations}
\label{eq:Eig_one}
\begin{alignat}{4}
\omega_{3, 1} &= \omega_0 \pm \sqrt{2} g, \\
      \omega_2 &= \omega_0,
\end{alignat}
\end{subequations}
revealing a solitary eigenfrequency $\omega_2$, which is unshifted from the bare resonance $\omega_0$, while the two other eigenfrequencies $\omega_3$ and $\omega_1$ display a splitting of $\sqrt{2} g$ from the central resonance. Incoherent processes in the trimer chain are taken into account via the quantum master equation~\cite{Gardiner2014,DowningValle2019,DowningValle2020}
\begin{equation}
\label{eq:master}
 \partial_t \rho = \mathrm{i} [ \rho, \hat{H} ] 
+ \sum_{ n = 1, 2, 3 } \frac{\gamma_{n}}{2} \mathcal{L} b_n \\
 + \sum_{ n = 1, 2, 3 } \frac{P_n}{2} \mathcal{L}^{\dagger} b_n,
\end{equation}
in terms of the Lindblad superoperators
\begin{subequations}
\label{eq:master2}
\begin{alignat}{3}
 \mathcal{L} b_n &= 2 b_n \rho b_n^{\dagger} -  b_n^{\dagger} b_n \rho - \rho b_n^{\dagger} b_n, \\
  \mathcal{L}^{\dagger} b_n &= 2 b_n^{\dagger} \rho b_n -  b_n b_n^{\dagger} \rho - \rho b_n b_n^{\dagger}.
 \end{alignat}
\end{subequations}
The unitary evolution is supplied by the commutator term on the right-hand-side of Eq.~\eqref{eq:master}, where the Hamiltonian operator $\hat{H}$ is given by Eq.~\eqref{eq:Ham}. Losses into the heat bath are tracked using the first Lindbladian term, where $\gamma_n \ge 0$ is the damping decay rate of the $n$th oscillator. Incoherent gain processes, where $P_n \ge 0$ is the pumping rate into oscillator $n$, are modeled by the final term in Eq.~\eqref{eq:master}.

In order to probe the mean-field dynamics, we exploit the property $\expect{\mathcal{O}} = \Tr{ \left( \mathcal{O} \rho \right) }$, for any operator $\mathcal{O}$, with the master equation of Eq.~\eqref{eq:master}. This procedure leads to the following Schr\"{o}dinger-like equation for the first moments of the trimer chain
\begin{equation}
\label{eqapp:of_motion_io}
\mathrm{i} \partial_t \psi = \mathcal{H} \psi,
\end{equation}
with the three-dimensional Bloch vector
\begin{equation}
\label{eqapp:of_motion_iwewo}
\psi =
\begin{pmatrix}
  \langle b_1 \rangle  \\
  \langle b_2 \rangle  \\
  \langle b_3 \rangle 
 \end{pmatrix},
\end{equation}
and where the dynamical matrix $\mathcal{H}$ of first moments reads
\begin{equation}
\label{eqapp:of_motion_io_2}
\mathcal{H} =
\begin{pmatrix}
\omega_{0} - \mathrm{i} \frac{\Gamma_1}{2} && g && 0  \\ 
g  && \omega_{0} - \mathrm{i} \frac{\Gamma_2}{2} && g\\ 
0 && g  && \omega_{0} - \mathrm{i} \frac{\Gamma_3}{2}
 \end{pmatrix}.
\end{equation}
In Eq.~\eqref{eqapp:of_motion_io_2}, the renormalized damping decay rate $\Gamma_n$ of each oscillator, due to the incoherent pumping $P_n$, is
\begin{equation}
\label{eqapp:ffsdfsg22}
\Gamma_n = \gamma_n - P_n.
\end{equation}
Let us consider the configuration where the first oscillator is subject to gain via $P_1 = \kappa$ (with $P_2 = P_3 = 0$), and the final oscillator can be described by the loss $\gamma_3 = \kappa$ (with $\gamma_1 = \gamma_2 = 0$). Then the mean-field theory of Eq.~\eqref{eqapp:of_motion_io_2} implies the $\mathcal{PT}$-symmetric Hamiltonian
\begin{align}
\label{eq:Ham_PT}
\hat{H}' =& \left( \omega_0 + \mathrm{i} \tfrac{\kappa}{2} \right) b_{1}^{\dagger} b_{1} + \omega_0 b_{2}^{\dagger} b_{2} + \left( \omega_0 - \mathrm{i} \tfrac{\kappa}{2} \right) b_{3}^{\dagger} b_{3} \nonumber \\
&+g \Big( b_{1}^{\dagger} b_{2} +  b_{2}^{\dagger} b_{3} + \mathrm{h.c.} \Big),
\end{align}
which is invariant under the necessary twin transformations of time and space, essentially that means $\mathrm{i} \to -\mathrm{i}$ and $(1,3) \to (3, 1)$. This $\mathcal{PT}$-symmetric arrangement of the trimer  chain is sketched in Fig.~\ref{exceptional}~(a), which highlights that the first oscillator is subject to gain $\kappa$ (yellow arrow), and the final oscillator to an equivalent loss $\kappa$ (purple arrow). The three eigenfrequencies $\omega_n'$ from Eq.~\eqref{eq:Ham_PT} are 
\begin{subequations}
\label{eq:Eig_one_PT}
\begin{alignat}{3}
\omega_{3, 1}' &= \omega_0  \pm  \Omega, \\
\omega_{2}' &= \omega_0,
\end{alignat}
\end{subequations}
where we have introduce the frequency
 \begin{equation}
\label{eqapp:ffsdfdfdfdsdsdfsg22}
 \Omega = \tfrac{1}{2} \sqrt{ 8 g^2 - \kappa^2 }.
 \end{equation}
Equation~\eqref{eq:Eig_one_PT} reveals the renormalization of the upper and lower eigenfrequency splittings from Eq.~\eqref{eq:Eig_one}, due to the incoherent processes associated with $\kappa$. In particular, there is an exceptional point [cf. Eq.~\eqref{eqapp:ffsdssdsddsdfsg22} for the dimer] at
 \begin{equation}
\label{eqapp:ffsdsdsdfsg22}
 \left( \frac{g}{\kappa} \right)_{\mathcal{N} = 3} = \frac{1}{2\sqrt{2}} \simeq 0.353...
 \end{equation}
which marks the border between the regime when the $\mathcal{PT}$ Hamiltonian is in its unbroken phase with real eigenvalues, $g \ge \kappa/(2\sqrt{2})$, and the broken phase with complex eigenvalues, $g < \kappa/(2\sqrt{2})$.

 \begin{figure}[tb]
 \includegraphics[width=0.75\linewidth]{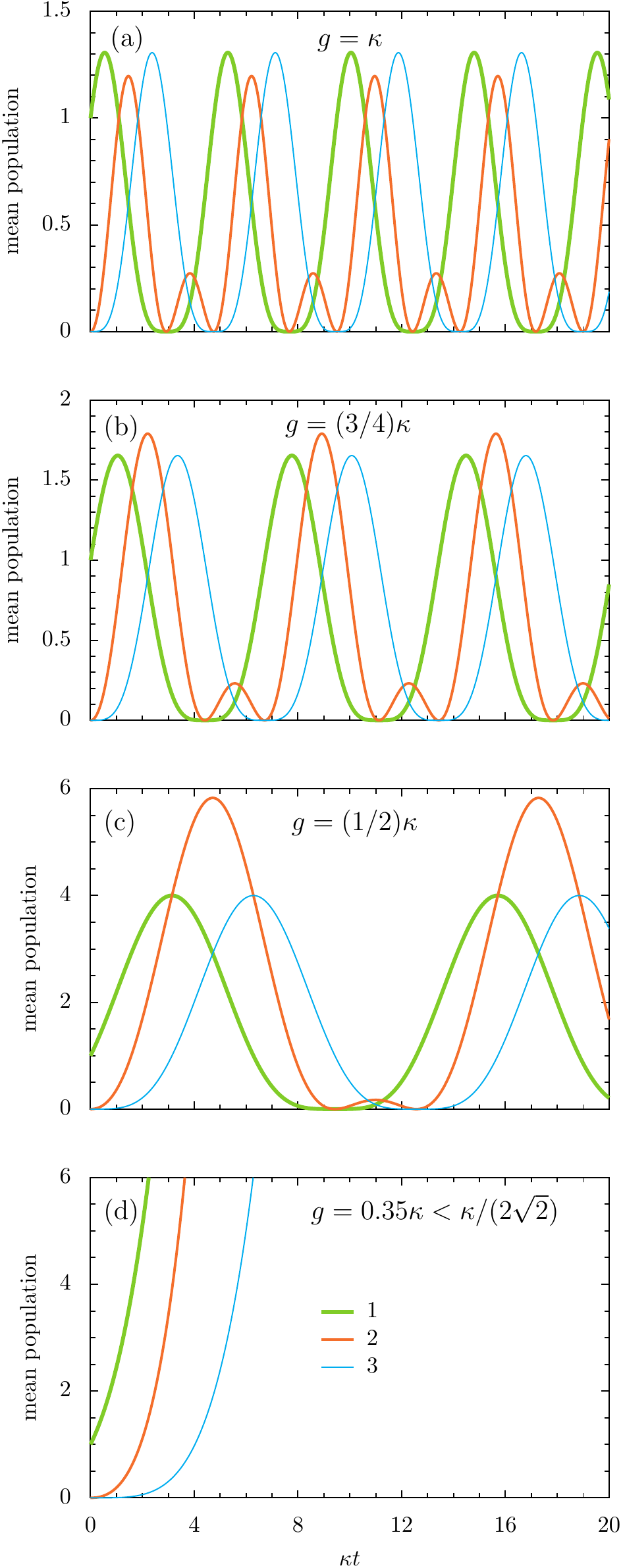}
 \caption{ \textbf{Population dynamics.} Evolution of the mean populations $\langle b_n^{\dagger} b_n \rangle$ along the trimer chain, as a function of time $t$, in units of the inverse loss-gain parameter $\kappa^{-1}$ [Eq.~\eqref{eq:magvhvhhster2}]. The coupling constant $g$ reduces from above to below the exceptional point upon descending the column of panels [Eq.~\eqref{eqapp:ffsdsdsdfsg22}]. The results for the first, second and third oscillators are denoted by the thick green, medium orange and thin cyan lines respectively [see the legend in panel (d), which applies to the whole figure].  }
 \label{dynam}
\end{figure}

We plot the eigenfrequencies of $\omega_{n}'$ in Fig.~\ref{exceptional} using Eq.~\eqref{eq:Eig_one_PT}, where real parts are given in panel (b) and the imaginary parts in panel (c). The exceptional point of Eq.~\eqref{eqapp:ffsdsdsdfsg22} is denoted by the dashed gray line, making explicit the broken and unbroken phases of the system. There are several features of Fig.~\ref{exceptional} which are shared amongst all odd-sited oligomers, namely: the purely real resonance frequency $\omega_0$ (orange line) is always an eigenfrequency, two eigenfrequencies always become complex in the broken $\mathcal{PT}$-symmetric phase, and these two aforementioned eigenfrequencies are always the two eigenfrequencies closest to $\omega_0$ (neglecting the trivial $\omega_0$ solution). Under the popular classification where an $n$-th order exceptional point refers to when $n$ eigenvalues coalesce at the exceptional point~\cite{Hodaei2017}, Fig.~\ref{exceptional}~(b, c) exposes a higher order exceptional point of the 3rd order (compared to 2nd order for a dimer, see Supplementary Note 1). These remarks are further justified in Supplementary Note 2, where analogous behavior with the quadrimer chain ($\mathcal{N} = 4$) is analyzed in detail, and some features associated with all even-sited oligomers are discussed in Supplementary Note 3.\\
 
\noindent \textbf{\textit{Trimer chain: dynamics}}\\
The equation of motion for the second moments of the trimer gives access to the mean populations $\langle b_n^{\dagger} b_n \rangle$ along the trimer chain. Similar to the calculation leading to Eq.~\eqref{eqapp:of_motion_io}, we obtain
\begin{equation}
\label{eqapp:of_motion}
\frac{\mathrm{d}}{\mathrm{d} t} \mathbf{u}  =  \mathbf{P} - \mathbf{M} \mathbf{u},
\end{equation}
for the 9-vector of correlators $\mathbf{u}$ and drive term $\mathbf{P}$, where
\begin{equation}
\label{eqapp:umatrix}
\mathbf{u} =
\begin{pmatrix}
  \mathbf{u}_1  \\
  \mathbf{u}_2  \\
  \mathbf{u}_2^{\dagger}
 \end{pmatrix}, \quad
  \mathbf{P} = 
\begin{pmatrix}
  P_1  \\
  P_2 \\
  P_3 \\
  ~\boldsymbol{0}_{6}
 \end{pmatrix},
 \end{equation}
where $\text{\bf{0}}_{n}$ is the zero matrix (of $n$-rows and a single column), and with the sub-vectors of $\mathbf{u}$
\begin{equation}
\label{eqapp:umatrixsdsdsd}
 \mathbf{u}_1 =
\left(\begin{array}{c}
  \langle b_1^{\dagger} b_1 \rangle  \\
  \langle b_2^{\dagger} b_2 \rangle  \\
  \langle b_3^{\dagger} b_3 \rangle  
    \end{array}
\right), \quad 
 \mathbf{u}_2 =
\left(\begin{array}{c}
 \langle b_1^{\dagger}  b_2 \rangle  \\
  \langle b_2^{\dagger}  b_3 \rangle   \\
  \langle b_3^{\dagger}  b_1 \rangle  \\
    \end{array}
\right).
 \end{equation} 
The matrix $\mathbf{M}$ of second moments in Eq.~\eqref{eqapp:of_motion} reads
\begin{equation}
\label{eqapp:Pdrive}
 \quad
\mathbf{M} = \begin{pmatrix}
  \mathbf{M}_{11} && \mathbf{M}_{12} && \mathbf{M}_{12}^{\ast}   \\
  \mathbf{M}_{12}^{\mathrm{T}} && \mathbf{M}_{22} && \mathbf{M}_{23}    \\
  \mathbf{M}_{12}^{\dagger} &&  \mathbf{M}_{23}^\ast && \mathbf{M}_{22}  
 \end{pmatrix},
 \end{equation}
 where the on-diagonal sub-matrices comprising $\mathbf{M}$ are 
 \begin{subequations}
\label{eqapp:umatrasasix22}
\begin{alignat}{3}
\mathbf{M}_{11} &= \mathrm{diag} \left( \Gamma_1, \Gamma_2, \Gamma_3 \right), \\
 \mathbf{M}_{22} &= \mathrm{diag} \left( \tfrac{\Gamma_1+\Gamma_2}{2}, \tfrac{\Gamma_2+\Gamma_3}{2}, \tfrac{\Gamma_3+\Gamma_1}{2} \right),
 \end{alignat}
 \end{subequations}
where $\Gamma_n$ is defined in Eq.~\eqref{eqapp:ffsdfsg22}, while the two off-diagonal sub-matrices of $\mathbf{M}$ are defined by
\begin{equation}
\label{eqapp:umatrix22}
\mathbf{M}_{12} =
\begin{pmatrix}
   \mathrm{i} g && 0 && 0 \\
   -\mathrm{i} g && \mathrm{i} g && 0 \\
    0 && -\mathrm{i} g &&  0
 \end{pmatrix},~
 \mathbf{M}_{23} =
\begin{pmatrix}
  0 && 0 && \mathrm{i} g \\
  0 && 0 &&  -\mathrm{i} g \\
   - \mathrm{i} g && \mathrm{i} g &&  0
 \end{pmatrix}.
\end{equation}
In Eq.~\eqref{eqapp:Pdrive}, the symbols $\ast, \dagger$ and $\mathrm{T}$ represent taking the conjugate, conjugate transpose, and transpose respectively.

Let us consider the $\mathcal{PT}$-symmetric arrangement of the trimer, as sketched in Fig.~\ref{exceptional}~(a). In this special configuration, the nontrivial eigenvalues of the matrix $\mathbf{M}$ in Eq.~\eqref{eqapp:Pdrive} are $\pm \mathrm{i} \sqrt{ 8 g^2 - \kappa^2}$ and $\pm \mathrm{i} \sqrt{ 8 g^2 - \kappa^2}/2$, recovering the criticality first expounded at the level of the non-Hermitian Hamiltonian in Eq.~\eqref{eqapp:ffsdsdsdfsg22}. Using the frequency $\Omega$ as defined in Eq.~\eqref{eqapp:ffsdfdfdfdsdsdfsg22}, we find the following analytic expressions for the populations
\begin{widetext}
\begin{subequations}
\label{eq:magvhvhhster2}
\begin{alignat}{3}
 \langle b_1^{\dagger} b_1 \rangle &= \frac{48 g^4 \Omega + 32 \kappa g^2 \Omega^2 \sin \left( \Omega t \right) + \kappa \left( \kappa^4 - 12 g^2 \kappa^2 + 32 g^4 \right) \sin \left(2 \Omega t \right) + 16 g^2 \Omega \left( 4 g^2 - \kappa^2 \right) \cos \left( \Omega t \right) +  \tilde{\Omega}^5 \cos \left( 2 \Omega t \right)}{32\Omega^5}, \\
 \langle b_2^{\dagger} b_2 \rangle &= \frac{g^2}{2\Omega^5} \sin^2 \left( \frac{\Omega t}{2} \right) \bigg\{ 8 g^2 \Omega + 4 \kappa \Omega^2 \sin \left( \Omega t \right) +2 \Omega \left( 4 g^2 - \kappa^2 \right) \cos \left( \Omega t \right) \bigg\}, \\
  \langle b_3^{\dagger} b_3 \rangle &= \frac{4 g^4}{\Omega^4} \sin \left( \frac{\Omega t}{2} \right), 
 \end{alignat}
\end{subequations}
\end{widetext}
where $\tilde{\Omega}^5 = 2 \Omega ( 8 g^4 - 8 g^2 \kappa^2 + \kappa^4 )$. Upon approaching the exceptional point (where $\Omega \to 0$), Eq.~\eqref{eq:magvhvhhster2} reduces to the algebraically divergent $\langle b_1^{\dagger} b_1 \rangle = (1+\kappa t/4)^4$, $\langle b_2^{\dagger} b_2 \rangle = (\kappa t/2)^2 (1+\kappa t/4)^2/2$, and $\langle b_3^{\dagger} b_3 \rangle = (\kappa t/4)^4$. Below the exceptional point, the trigonometric functions in Eq.~\eqref{eq:magvhvhhster2} are superseded by hyperbolic functions, leading to exponential divergencies.

 \begin{figure}[tb]
 \includegraphics[width=1.0\linewidth]{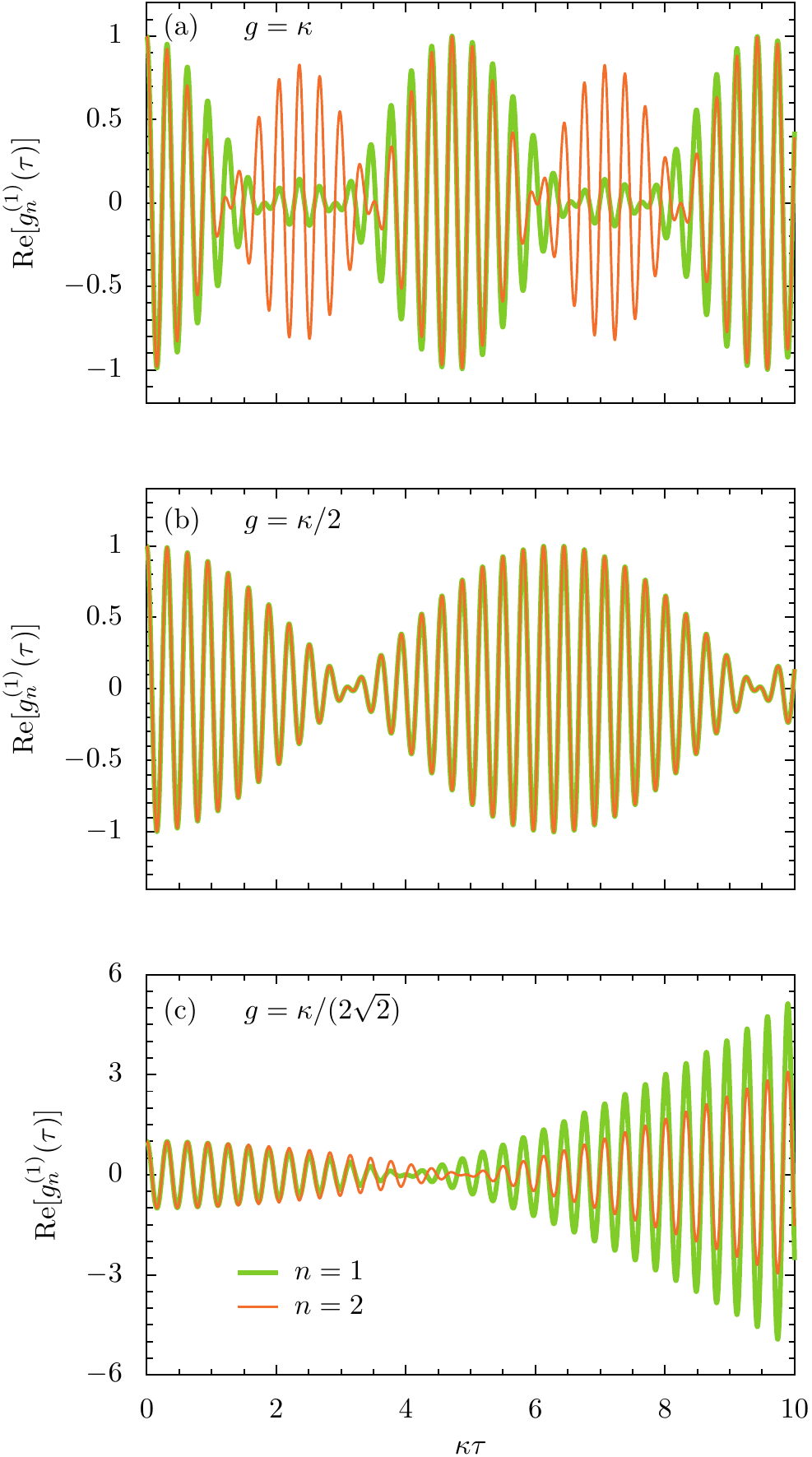}
 \caption{ \textbf{Dynamics of the correlations.} Evolution of the real part of the first-order correlation function $g^{(1)}(\tau)$, as a function of the time delay $\tau$, in units of the inverse loss-gain parameter $\kappa^{-1}$ [Eq.~\eqref{eqapp:umasdsdcvsdscvaasdadsatrix}]. Panel (a, b, c): the coupling constant $g$ is reduced from above (upper two panels) to at (lower panel) the exceptional point located at $g = \kappa/(2\sqrt{2})$. The results for the first and second oscillators are denoted by the thick green and thin orange lines respectively [see the legend in panel (c), which applies to the whole figure]. In the figure, $\omega_0 = 20 \kappa$.}
 \label{corr}
\end{figure}

We plot the populations $\langle b_n^{\dagger} b_n \rangle$ of the first, second and third oscillators ($n = 1, 2, 3$) as the thick green, medium orange and thin cyan lines in Fig.~\ref{dynam}, using the solutions of Eq.~\eqref{eq:magvhvhhster2}. In Fig.~\ref{dynam}~(a), where $g = \kappa$, a high frequency population cycle is observed, which is maintained over time due to the balanced loss and gain in the system. In panels (b) and (c), where the coupling strength is reduced to $g = 3\kappa/4$ and $g = \kappa/2$ respectively, the frequency of the population cycle is successively reduced, while the maxima of the mean populations are increased due to the closening proximity to the exceptional point [cf. Eq.~\eqref{eqapp:ffsdsdsdfsg22}]. The broken $\mathcal{PT}$ phase is exemplified in panel (d), where $g = 0.35 \kappa < \kappa/(2\sqrt{2})$, which displays the characteristically diverging population dynamics associated with breakdown beyond the exceptional point.\\

\noindent \textbf{\textit{Trimer chain: correlations}}\\
The temporal coherence can be quantified using the first-order correlation function~\cite{Gardiner2014}
\begin{equation}
\label{eqapp:umsdsdasdsdaasdadsatrix}
g_n^{(1)} (\tau)  =  \lim_{t \to \infty} \frac{\langle b_n^{\dagger} (t) b_n (t+\tau) \rangle}{\langle b_n^{\dagger} (t) b_n (t) \rangle},
 \end{equation}
 where $\tau$ is the time delay, and where the normalization is taken over a long time scale $t \to \infty$. This quantity has the property that perfect coherence is associated with $|g_n^{(1)} (\tau)|  = 1$ and complete incoherence corresponds to $|g_n^{(1)} (\tau)|  = 0$, while intermediate cases specify the degree of partial coherence. The manipulations resulting in Eq.~\eqref{eqapp:of_motion_io}, and an application of the quantum regression theorem, lead to an equation for the first desired two-time correlator $\langle b_1^{\dagger} (t) b_1 (t+\tau) \rangle$, via
\begin{equation}
\label{eqapp:umsdsdasdadsatrix}
\partial_{\tau} \mathbf{v}  + \mathbf{Q} \mathbf{v} = 0, \quad\quad
\mathbf{v} =
\begin{pmatrix}
  \langle b_1^{\dagger} (t) b_1  (t+\tau) \rangle \\
  \langle b_1^{\dagger} (t) b_2  (t+\tau) \rangle \\
    \langle b_1^{\dagger} (t) b_3  (t+\tau) \rangle \\
 \end{pmatrix},  \quad\quad
 \end{equation}
with the $3 \times 3$ regression matrix
\begin{equation}
\label{eqapp:umsdsdadssdasdadsatrix}
\mathbf{Q} = \begin{pmatrix}
 \mathrm{i} \omega_0 + \frac{\Gamma_1}{2} &&  \mathrm{i} g && 0  \\
\mathrm{i} g &&  \mathrm{i} \omega_0 + \frac{\Gamma_2}{2} && \mathrm{i} g  \\
0 &&  \mathrm{i} g &&   \mathrm{i} \omega_0 + \frac{\Gamma_3}{2}  
 \end{pmatrix}.
 \end{equation}
Similar equations may be derived for $\langle b_2^{\dagger} (t) b_2 (t+\tau) \rangle$ and $\langle b_3^{\dagger} (t) b_3 (t+\tau) \rangle$. The solution of Eq.~\eqref{eqapp:umsdsdasdadsatrix}, along with the definition of Eq.~\eqref{eqapp:umsdsdasdsdaasdadsatrix}, leads to the sought after first-order correlation functions. In the $\mathcal{PT}$-symmetric setup of trimer, as drawn in Fig.~\ref{exceptional}~(a), one finds the neat expressions
 \begin{subequations}
\label{eqapp:umasdsdcvsdscvaasdadsatrix}
\begin{alignat}{3}
g_{1}^{(1)} (\tau)  &= g_{3}^{(1)} (\tau)  = \frac{  8 g^2 \cos^2 \left( \frac{\Omega \tau}{2} \right) - \kappa^2 }{ 4\Omega^2  }  \mathrm{e}^{-\mathrm{i} \omega_0 \tau},  \\
g_2^{(1)} (\tau)  &=  \frac{ 4 g^2 \kappa^2 - \kappa^4 + 32 g^4 \cos \left( \Omega \tau \right)  }{ 4 g^2 \kappa^2 - \kappa^4 + 32 g^4  }  \mathrm{e}^{-\mathrm{i} \omega_0 \tau},  
\end{alignat}
 \end{subequations}
 which characteristically include the harmonic component $\mathrm{e}^{-\mathrm{i} \omega_0 \tau}$, representing a monochromatic field centered on $\omega_0$, and a pre-factor accounting for the specific $\mathcal{PT}$-symmetric setup of the trimer. In the limit of $\Omega \to 0$, that is approaching the exceptional point $g \to \kappa/(2\sqrt{2})$, Eq.~\eqref{eqapp:umasdsdcvsdscvaasdadsatrix} tends towards the quadratically divergent results $g_{1, 3}^{(1)} (\tau) \to \{ 1 - \kappa^2 \tau^2 /16 \} \mathrm{e}^{-\mathrm{i} \omega_0 \tau}$ and $g_{2}^{(1)} (\tau) \to \{ 1 - \kappa^2 \tau^2/24 \} \mathrm{e}^{-\mathrm{i} \omega_0 \tau}$. For coupling strengths below the exceptional point the trigonometric functions are replaced with hyperbolic functions, indicating exponentially divergent behavior. We plot the real parts of the coherences $g_{1}^{(1)} (\tau)$ and $g_{2}^{(1)} (\tau)$ of the first and second oscillators as the thick green and thin orange lines in Fig.~\ref{corr}, using the solutions of Eq.~\eqref{eqapp:umasdsdcvsdscvaasdadsatrix}. In Fig.~\ref{corr}~(a), well above the exceptional point at $g = \kappa$, the $\mathcal{PT}$ symmetry ensures an undamped periodic response, with rapid oscillations and a dynamic behaviour satisfying $0 < |g_n^{(1)} (\tau)|  < 1$. Exactly at $g = \kappa/2$, where all three coherences are accidentally equal as shown in panel~(b), a well-defined wave envelope develops. In panel (c), at the exceptional point $g = \kappa/(2\sqrt{2})$, there is initially regular, high frequency oscillations due to short time behaviour being essentially dominated by the zeroth order term $g_{n}^{(1)} (\tau) \simeq \mathrm{e}^{-\mathrm{i} \omega_0 \tau}$. Once the quadratic correction in $\kappa t$ becomes non-negligible, the divergence characteristic of the broken $\mathcal{PT}$ symmetric phase finally emerges.\\

\begin{figure}[tb]
 \includegraphics[width=1.0\linewidth]{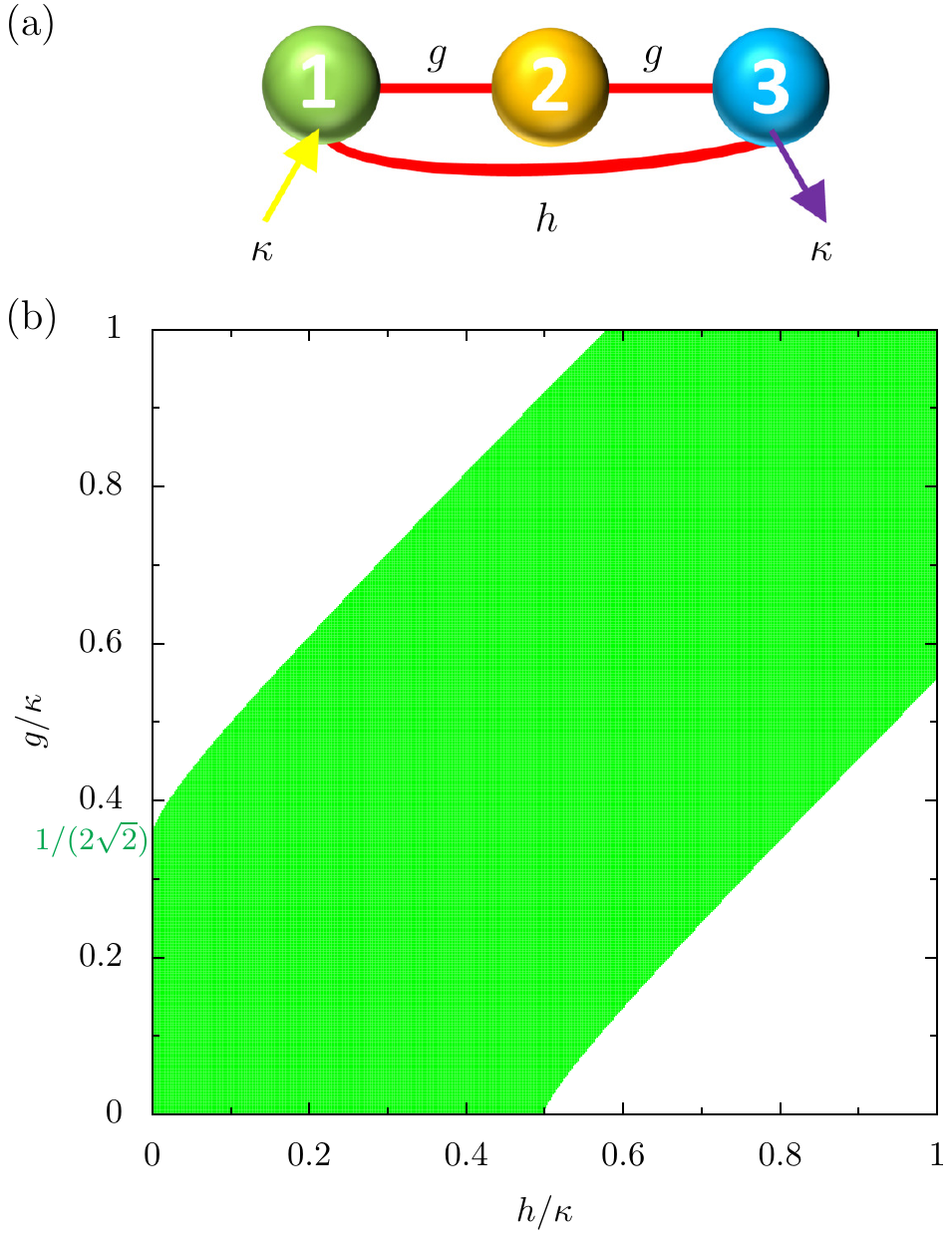}
 \caption{ \textbf{The effect of long-range interactions.} Panel (a): a sketch of the $\mathcal{PT}$-symmetric trimer (colored spheres) beyond nearest-neighbor coupling, where the first-neighbor coupling constant is $g$, and the second-neighbor coupling constant is $h$. The first oscillator (green sphere) is subject to gain $\kappa$ (yellow arrow), and the final oscillator (cyan sphere) to loss $\kappa$ (purple arrow). Panel (b): the $\mathcal{PT}$ symmetry phase diagram of the trimer, given by the evolution of the exceptional point $(g/\kappa)_3$ with the first and second-neighbor couplings $g$ and $h$, both in units of $\kappa$ [Eq.~\eqref{eq:Eig_one_PsdsT}]. White: unbroken phase. Green: broken phase. }
 \label{beyond}
\end{figure}

\begin{figure*}[tb]
 \includegraphics[width=1.0\linewidth]{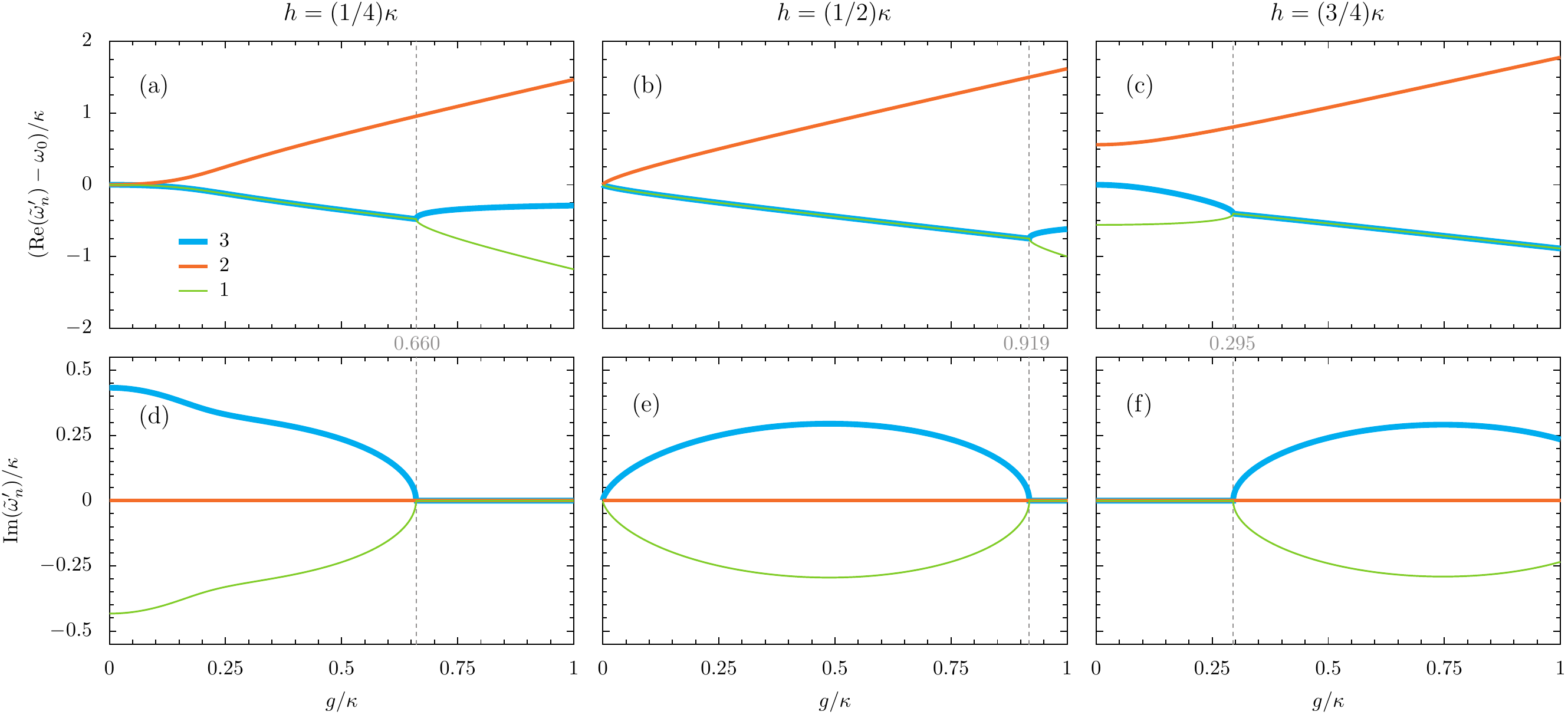}
 \caption{ \textbf{Exceptional points of the trimer with long-range interactions.}  Panels (a-c): the real parts of the eigenfrequencies $\tilde{\omega}_n'$ of the trimer, measured from $\omega_0$, as a function of $g$, in units of $\kappa$ [Eq.~\eqref{eq:Eig_one_PsdsT}]. Panels (d-f): the corresponding imaginary parts of the eigenfrequencies $\tilde{\omega}_n'$. Vertical dashed lines: exceptional points marking broken-unbroken $\mathcal{PT}$ symmetry phases. In the first, second, and third columns, the second-nearest neighbor coupling constant $h = \kappa/4, \kappa/2$ and $3\kappa/4$ respectively. The results for the first, second and third eigenfrequencies $\tilde{\omega}_n'$ are denoted by the thin green, medium orange and thick cyan lines respectively [see the legend in panel (a), which applies to the whole figure].}
 \label{TrimerNN}
\end{figure*}

\noindent \textbf{\textit{Trimer chain: long-range coupling}}\\
Let us now consider the effects of going beyond the nearest-neighbor coupling approximation employed in Eq.~\eqref{eq:Ham}. To do so, we introduce the second-nearest neighbor coupling constant $h$, which connects the first and third oscillators, via the generalized Hamiltonian $\hat{H}' = \hat{H} + h ( b_{1}^{\dagger} b_{3} + b_{3}^{\dagger} b_{1} )$, where $\hat{H}$ is defined in Eq.~\eqref{eq:Ham}. This extension leads to a generalization of the eigenfrequencies of Eq.~\eqref{eq:Eig_one} to $\tilde{\omega}_{n}$, where
\begin{subequations}
\label{eq:Eig_onssdsdsde}
\begin{alignat}{4}
\tilde{\omega}_{3, 1} &= \omega_0 + \tfrac{h}{2} \pm \tfrac{1}{2} \sqrt{8 g^2 + h^2}, \\
      \tilde{\omega_2} &= \omega_0 - h.
\end{alignat}
\end{subequations}
The associated $\mathcal{PT}$-symmetric setup of trimer chain is sketched in Fig.~\ref{beyond}~(a) [cf. Fig.~\ref{exceptional}~(a)]. The resulting eigenfrequencies $\tilde{\omega}_{n}'$ are [cf. Eq.~\eqref{eq:Eig_one_PT}]
\begin{subequations}
\label{eq:Eig_one_PsdsT}
\begin{alignat}{3}
\tilde{\omega}_{3}' &= \omega_0 + \sqrt{\frac{8 g^2 + 4 h^2 - \kappa^2 }{3}} \cos \left( \frac{\alpha}{3} \right), \\
\tilde{\omega}_{2}' &= \omega_0 + \sqrt{\frac{8 g^2 + 4 h^2 - \kappa^2 }{3}} \cos \left( \frac{\alpha + 4 \pi}{3} \right), \\
\tilde{\omega}_{1}' &= \omega_0 + \sqrt{\frac{8 g^2 + 4 h^2 - \kappa^2 }{3}} \cos \left( \frac{\alpha + 2 \pi}{3} \right), 
\end{alignat}
\end{subequations}
where we have introduced the quantity
 \begin{equation}
\label{eqapp:sdsdsdsds22}
 \alpha = \arccos \left( \frac{ 24 \sqrt{3} g^2 h }{ \left( 8 g^2 + 4 h^2 - \kappa^2  \right)^{3/2} } \right).
 \end{equation}
The inclusion of second-nearest neighbor coupling $h$ leads to a significantly richer phase diagram than with nearest-neighbor coupling only, as is demonstrated in Fig.~\ref{beyond}~(b). Notably, when $h = 0$ Eq.~\eqref{eqapp:ffsdsdsdfsg22} is recovered, so that above this threshold strength of $1/(2\sqrt{2})$ the system is in its unbroken  phase. With increasing $h$, the exceptional point $(g/\kappa)_3$ increases in value, up until $h = \kappa/2$. Above this critical point, the unbroken phase can be explored either with weak enough $g$, or strong enough $g$, with a region of broken phase in between. This causes a green stripe in the phase diagram of Fig.~\ref{beyond}~(b), which notably contains the equal coupling ($h = g$) ring-like limit. The aforementioned broken-unbroken transitions from above and from below can be explicitly seen in Fig.~\ref{TrimerNN}, where the real and imaginary parts of $\tilde{\omega}_{n}'$ are shown, as a function of $g/\kappa$, in the upper and lower rows respectively. In the first column of Fig.~\ref{TrimerNN}, one notices how a nonzero second-nearest neighbor coupling ($h = \kappa/4$) has led to a larger exceptional point of $(g/\kappa)_3 \simeq 0.660$, compared to the nearest-neighbor coupling case when $(g/\kappa)_3 \simeq 0.353$. The middle column, at the critical point of $h = \kappa/2$, shows the onset of a new region of unbroken $\mathcal{PT}$ phase for vanishingly small $g$. This novel region is even more apparent in the final column of Fig.~\ref{TrimerNN}, where $h = 3\kappa/4$ and the exceptional point is well below the nearest-neighbor value, being $(g/\kappa)_3 \simeq 0.295$. Across all of these cases, it is most apparent that the higher (3rd order) exceptional point of the trimer with nearest-neighbor coupling only [cf. Fig.~\ref{exceptional}~(b, c)] has been downgraded to a standard 2nd order exceptional point in Fig.~\ref{TrimerNN}. This is due to the long-range interactions perturbing the eigensolution otherwise residing exactly at $\omega_0$. Similarly rich features due to long-range interactions are also seen in the quadrimer chain ($\mathcal{N} = 4$), as is demonstrated in Supplementary Note 2.

%===========================================================================
%===========================================================================
%===========================================================================
%===========================================================================
%===========================================================================
%===========================================================================
%===========================================================================
%===========================================================================

\begin{figure}[tb]
 \includegraphics[width=1.0\linewidth]{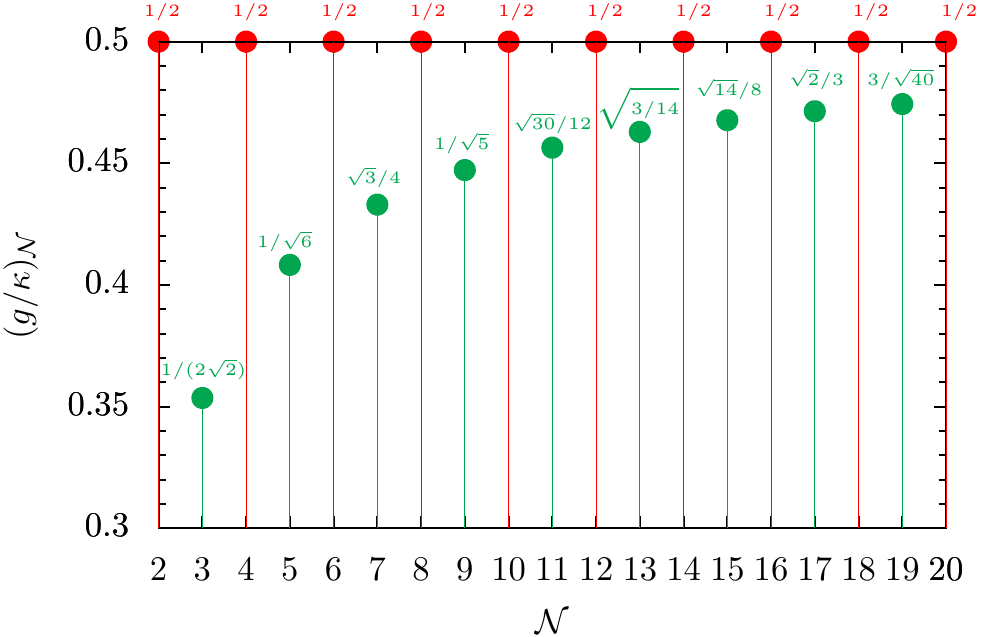}
 \caption{ \textbf{Evolution of the exceptional point in arbitrary chains.}  The exceptional point $(g/\kappa)_\mathcal{N}$ as a function of the number of sites $\mathcal{N}$ in the oligomer chain [Eq.~\eqref{eqapp:sdsdsddss22}]. Results with even (odd) $\mathcal{N}$ are associated with red (green) circles.}
 \label{critical}
\end{figure}

 \noindent \textbf{\textit{Oligomer chains}}\\
We have seen some fundamental properties of short $\mathcal{PT}$-symmetric oligomer chains (specifically for $\mathcal{N} = 3$, and for $\mathcal{N} = 2$ and $\mathcal{N} = 4$ in the Supplementary Notes 1 and 2). Let us now consider a general oligomer of arbitrary size $\mathcal{N}$, with nearest-neighbor coupling only. The eigenfrequencies read
 \begin{equation}
\label{eqapp:sdsds22}
 \omega_{n}(\mathcal{N}) = \omega_0 + 2 g \cos \left( \frac{n \pi}{\mathcal{N}+1} \right),
 \end{equation}
where the index $n \in [1, \mathcal{N}]$ labels each mode [such that the specific results for $\mathcal{N} = 3$ reproduce Eq.~\eqref{eq:Eig_one}]. If we generalize the $\mathcal{PT}$-symmetric arrangement of Fig.~\ref{exceptional}~(a), that is if we allow for gain into the first oscillator and an equivalent loss out of the $\mathcal{N}$th oscillator in the chain, so that the setup is $[\mathrm{gain}]-[\mathrm{neutral]_{\mathcal{N}-2}-[\mathrm{loss}}]$, we can find how the exceptional point $(g/\kappa)_\mathcal{N}$ evolves as a function of $\mathcal{N}$. The result of this diagonalization of a chain of an arbitrary $\mathcal{N}$-oscillator oligomer is derived in Supplementary Note 3, using a transfer matrix method~\cite{Saroka2017, Saroka2018, Klein1994}. This procedure leads to the rather beautiful expression [cf. Eq.~\eqref{eqapp:ffsdssdsddsdfsg22} and Eq.~\eqref{eqapp:ffsdsdsdfsg22}]
 \begin{equation}
\label{eqapp:sdsdsddss22}
\left( \frac{g}{\kappa} \right)_{\mathcal{N}} = \begin{cases}
\frac{1}{2}, &\mbox{ } \mathcal{N} =  2,4,6,... \\
\frac{1}{2} \sqrt{\frac{\mathcal{N}-1}{\mathcal{N}+1}}, & \mbox{ } \mathcal{N} =  3,5,7,...
\end{cases}
 \end{equation}
We display graphically the formula of Eq.~\eqref{eqapp:sdsdsddss22} in Fig.~\ref{critical}. Most notably, for oligomers of even size $\mathcal{N}$ (red circles), the exceptional point is constant at $(g/\kappa)_\mathcal{N} = 1/2$, and is of 2nd order. However, oligomers of odd size $\mathcal{N}$ (green circles) are associated with smaller exceptional points than the celebrated dimer result, and are of higher (3rd) order. These exceptional points are bounded by the limiting cases of the trimer result of $(g/\kappa)_3  = 1/(2 \sqrt{2})\simeq 0.353...$ and the infinitely long chain result of $(g/\kappa)_\infty  = 1/2$, as shown in Fig.~\ref{critical} for chains up to $\mathcal{N}= 20$ oscillators. In particular, the even-odd behavior shown in Fig.~\ref{critical} is ripe for future experimental detection, as is the trend for increasing large values of the exceptional point with increasingly long odd-numbered chains, following the trend encapsulated by Eq.~\eqref{eqapp:sdsdsddss22}, and its inverse-linear asymptotics $(g/\kappa)_\mathcal{N}  \simeq (1-\mathcal{N}^{-1})/2$ for large $\mathcal{N}$. While we do not account for disorder, or for dimerization of the chain (which may be interesting from a topological point of view~\cite{Martinez2021}), such extensions can be readily taken care of within this framework.

\begin{figure*}[tb]
 \includegraphics[width=1.0\linewidth]{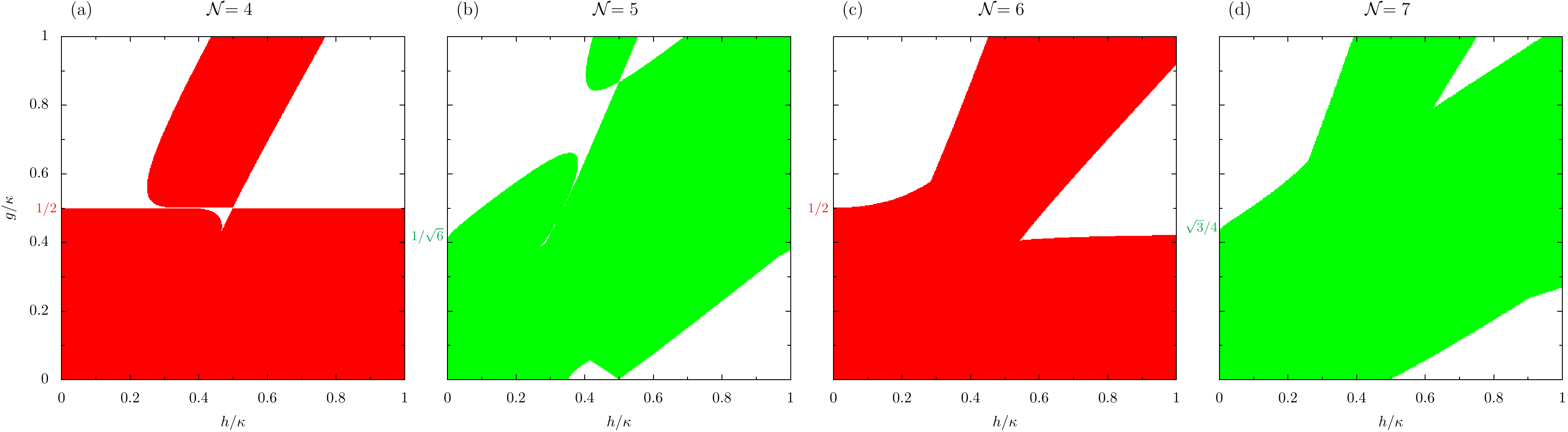}
 \caption{  \textbf{Phase diagrams of oligomer chains with long-range interactions.} Panels (a, b, c, d): $\mathcal{PT}$ symmetry phase diagrams of oligomer chains of size $\mathcal{N} = \{ 4, 5, 6, 7 \}$, given by the evolution of the exceptional point $(g/\kappa)_\mathcal{N}$, as a function of the first and second-neighbor couplings $g$ and $h$, both in units of $\kappa$. White: unbroken phase. Colors: broken phase. The analytic results in the $h \to 0$ limit are marked on the vertical axis [Eq.~\eqref{eqapp:sdsdsddss22}]. The $\mathcal{N} = 3$ case was already presented in Fig.~\ref{beyond}~(b). }
 \label{sev}
\end{figure*}

The addition of next-nearest neighbor hoppings to oligomers of an arbitrary length allows us to generalize our investigation of long-range interactions in a short trimer chain [cf. Fig.~\ref{beyond}~(b)]. Similar to the $\mathcal{N} = 3$ case, we can map the phase diagram marking the regions of broken (colored) and unbroken (white) $\mathcal{PT}$ symmetric phase, as is shown in Fig.~\ref{sev}~(a, b, c, d) for oligomers of length $\mathcal{N} = \{ 4, 5, 6, 7 \}$. The two relevant parameters are the first and second-neighbor coupling strengths $g$ and $h$, such that the vertical axis ($h = 0$) is marked with analytic results from Eq.~\eqref{eqapp:sdsdsddss22}. Away from this point, the influence of nonzero next-nearest neighbor hopping is rather profound: leading to seas (and even enclaves) of unbroken $\mathcal{PT}$ symmetry in a variety of geometries. Recent advances with so-called programmable interactions in atomic arrays suggest that the experimental exploration of such phase diagrams is increasingly accessible~\cite{Schleier2021}, aside from the demonstrated tunable interaction ranges in trapped atomic ions~\cite{Jurcevic2014, Manovitz2020}.\\ 

%===========================================================================
%===========================================================================
%===========================================================================
%===========================================================================
%===========================================================================
%===========================================================================
%===========================================================================
%===========================================================================

\noindent \textbf{Conclusions}\\
We have considered some fundamental properties of oligomers of an arbitrary size which satisfy $\mathcal{PT}$ symmetry due to having gain into the first oscillator and an equivalent loss out of the final oscillator. We have unveiled analytically the behavior of the exceptional points as a function of the chain length, which governs the stability of the population dynamics in the system and the presence of amplification. In particular, we have reported an even-odd effect for oligomers of increasing size, derived the bounds on all possible exceptional points, and mapped the relevant phase diagrams when long-range interactions are taken into account.

Focusing on short oligomers, we have provided simple quantum theories locating their exceptional points, and in doing so we found unconventional population dynamics and interesting first-order coherences near to the unbroken-broken $\mathcal{PT}$-symmetric phases. We have also discussed effects beyond nearest-neighbor coupling, which leads to rich $\mathcal{PT}$ symmetry phase diagrams. In particular, we have shown that reaching the unbroken $\mathcal{PT}$ symmetric phase is no longer purely dependent on going above a threshold value of coupling-to-dissipation strength $g/\kappa$, rather one may also go below a different threshold value, such that the broken phase can live in a sweet-spot in-between.

Our versatile theory is relevant across a number of optical and photonic platforms, including coupled ring resonators~\cite{Hodaei2014}, coupled cavities~\cite{Hodaei2016}, coupled waveguides~\cite{Feng2011,Xia2021} and meta-atoms~\cite{Sturges2020}. Our theoretical results provide a route-map for the scaling up of $\mathcal{PT}$-symmetric systems, and paves the way for the observation of cooperative effects in arbitrarily large systems. There are clear perspectives for the experimental detection of our predictions, including finite size effects, even-odd behaviors, unconventional light transport and correlations, and long-range interactions leading to sweet spot regions of $\mathcal{PT}$ symmetry phase breakdown.\\

%===========================================================================
%===========================================================================
%===========================================================================
%===========================================================================
%===========================================================================
%===========================================================================
%===========================================================================
%===========================================================================

\noindent \textbf{Methods}\\
In this theoretical work, the methods used are quantum master equations (as described in the main text [cf. Eq.~\eqref{eq:master}] and Supplementary Note 1), and an extended transfer matrices method (as detailed in Supplementary Note 3).\\

\noindent \textbf{Data availability}\\
Data sharing is not applicable to this article as no datasets were generated or analysed during the current theoretical study.\\

\noindent \textbf{Acknowledgments}\\
\textit{Funding}: CAD is supported by a Royal Society University Research Fellowship (URF\slash R1\slash 201158), and via the Royal Society Research Grant (RGS\slash R1 \slash 211220). VAS is supported by the Research Council of Norway Center of Excellence funding scheme (project no. 262633, ``QuSpin''). \textit{Discussions}: We are grateful to A.~Qaiumzadeh and T.~J.~Sturges for useful conversations.\\

\noindent \textbf{Author contributions}\\
CAD conceived of the study, performed the calculations with the quantum master equations, and wrote the first version of the manuscript. VAS worked on the transfer matrices method and redrafted the manuscript. Both authors gave final approval for publication and agree to be held accountable for the work performed therein.\\

\noindent \textbf{Competing interests}\\
 The authors declare no competing interests. They have no competing financial or non-financial interests as defined by Nature Research.\\

\noindent \textbf{Additional information}\\
Supplementary Information is available for this paper.\\

\noindent \textbf{ORCID}\\
CAD: \href{https://orcid.org/0000-0002-0058-9746}{0000-0002-0058-9746}. \\
VAS: \href{https://orcid.org/0000-0002-8980-6611}{0000-0002-8980-6611}.\\

%===========================================================================
%===========================================================================
%===========================================================================
%===========================================================================
%===========================================================================
%===========================================================================
%===========================================================================
%===========================================================================

\end{document}